# Nonadditive measure and quantum entanglement in a class of mixed states of $N^n$-system

Sumiyoshi Abe

*College of Science and Technology, Nihon University,*

*Funabashi, Chiba 274-8501, Japan*

Through the generalization of Khinchin's classical axiomatic foundation, a basis is developed for nonadditive information theory. The classical nonadditive conditional entropy indexed by the positive parameter $q$ is introduced and then translated into quantum information. This quantity is nonnegative for classically correlated states but can take negative values for entangled mixed states. This property is used to study quantum entanglement in the parametrized Werner-Popescu-like state of an $N^n$-system, that is, an $n$-partite $N$-level system. It is shown how the strongest limitation on validity of local realism (i.e., separability of the state) can be obtained in a novel manner.

PACS numbers: 03.65.Bz, 03.67.-a, 05.20.-y, 05.30.-d

# I. INTRODUCTION

There is growing interest in roles of nonadditive measures in quantum information theory. In Ref. [1], inadequacy of the additive Shannon-von Neumann entropy as a measure of the information content of a quantum system has been pointed out, for example. Also, there is a theoretical observation [2] that the measure of quantum entanglement may not be additive.

Recently, the explicit use of nonadditive quantum information measures has been made in quantum information theory [3-6]. These attempts are primarily based on the Tsallis entropy, which is defined as follows:

$$S_q[\hat{\rho}] = \frac{1}{1-q}\left(\text{Tr}\,\hat{\rho}^q - 1\right), \tag{1}$$

where $\hat{\rho}$ is the system density matrix and $q$ is the positive entropic index. This quantity is regarded as a one-parameter generalization of the von Neumann entropy, which is obtained in the limit $S_q[\hat{\rho}] \to S[\hat{\rho}] = -\text{Tr}(\hat{\rho}\ln\hat{\rho})$ $(q \to 1)$. $S_q[\hat{\rho}]$ possesses some important properties as an entropy. It is nonnegative, definitely concave for all values of $q > 0$, and fulfills the $H$-theorem. Additivity is, however, to be replaced by *pseudoadditivity*, which means that, for a product state of a bipartite system $(A, B)$, the total amount satisfies

$$S_q[\hat{\rho}(A) \otimes \hat{\rho}(B)] = S_q[\hat{\rho}(A)] + S_q[\hat{\rho}(B)] + (1-q)S_q[\hat{\rho}(A)]S_q[\hat{\rho}(B)]. \qquad (2)$$

Clearly, additivity holds only in the limit $q \to 1$. In recent years, $S_q[\hat{\rho}]$ and its classical counterpart have been widely discussed in the area of *nonextensive statistical mechanics* [7].

In this paper, we develop a basis for nonadditive quantum information theory and then apply it to the study of quantum entanglement in an $N^n$-system (i.e., an $n$-partite $N$-level system). The quantum states we consider here are described by a class of density matrices, which are the generalizations of the parametrized $2 \times 2$ Werner-Popescu state. We show how, in a novel manner, the strongest limitation on validity of local realism (i.e., separability of the density matrices) can be obtained for such a class of states by using the nonadditive conditional entropy.

## II. CLASSICAL NONADDITIVE MEASURE AND ITS AXIOMATIC FOUNDATION

Although our interest is in quantum theory, it seems appropriate to mention here that, at the classical level, the Tsallis entropy has its mathematical characterization, like the Shannon entropy [8,9]. Therefore, we wish to devote this section to a brief summary of this point.

The axioms and the uniqueness theorem have been presented for the Tsallis entropy in Ref. [10], in which the nonadditive conditional entropy has been introduced for the first time. Thereby, the Shannon-Khinchin axiomatic framework [8,9] was generalized to nonadditive information theory. The set of axioms presented in Ref. [10] is the following: [I] $S_q(p_1, p_2, \ldots, p_W)$ is continuous with respect to all its arguments and takes its maximum for the equiprobability distribution $p_i = 1/W$ $(i = 1, 2, \ldots, W)$, [II] $S_q(A, B) = S_q(A) + S_q(B|A) + (1-q) S_q(A) S_q(B|A)$ for a composite system $(A, B)$, and [III] $S_q(p_1, p_2, \ldots, p_W, p_{W+1} = 0) = S_q(p_1, p_2, \ldots, p_W)$. It can be shown [10] that the quantity $S_q$ satisfying [I]-[III] is, up to a multiplicative constant, uniquely given by

$$S_q(p_1, p_2, \ldots, p_W) \equiv S_q[p] = \frac{1}{1-q} \left[ \sum_{i=1}^{W} (p_i)^q - 1 \right], \tag{3}$$

which is the classical counterpart of $S_q[\hat{\rho}]$ in Eq. (1). Comparing this set of axioms with that of Khinchin [9], we see that the one and only difference is in [II], where $S_q(B|A)$ is the nonadditive conditional entropy defined as follows:

$$S_q(B|A) = \left\langle S_q(B|A_i) \right\rangle_q^{(A)}, \tag{4}$$

provided that $S_q(B|A_i)$ is the Tsallis entropy of the conditional probability distribution of $B$ with $A$ found in its $i$th state, $p_{ij}(B|A) = p_{ij}(A, B) / p_i(A)$ with the marginal probability distribution $p_i(A) = \sum_j p_{ij}(A, B)$. The symbol $\langle Q \rangle_q^{(A)}$ stands for the *normalized q-expectation value* [11] defined by

$$\langle Q \rangle_q^{(A)} = \sum_i Q_i P_i(A), \tag{5}$$

where

$$P_i(A) = \frac{[p_i(A)]^q}{\sum_i [p_i(A)]^q} \tag{6}$$

is the *escort distribution* associated with $p_i(A)$, which has originally been introduced in the context of statistical mechanical description of chaotic systems [12].

The generalized composition law in [II] can be ascertained by using Eqs. (3)-(6). From it, immediately follows another expression

$$S_q(B|A) = \frac{S_q(A, B) - S_q(A)}{1 + (1-q) S_q(A)}, \tag{7}$$

where $S_q(A)$ is the Tsallis entropy of the marginal probability distribution $p_i(A)$.

Note that there exists the following correspondence relation between the Bayes law and the generalized composition law in [II]:

$$p_{ij}(A, B) = p_i(A) p_{ij}(B|A) = p_j(B) p_{ij}(A|B)$$

$$\leftrightarrow$$

$$\begin{aligned} S_q(A, B) &= S_q(A) + S_q(B|A) + (1-q) S_q(A) S_q(B|A) \\ &= S_q(B) + S_q(A|B) + (1-q) S_q(B) S_q(A|B), \end{aligned} \qquad (8)$$

which is natural in view of pseudoadditivity. [See Eq. (2).]

Let us further discuss the generalized composition law for a multipartite system. To be specific, here we consider a tripartite system, $(A, B, C)$, as a simple example. The Bayes multiplication rule reads

$$p_{ijk}(A, B, C) = p_{jk}(B, C) p_{ijk}(A|B, C) = p_k(C) p_{jk}(B|C) p_{ijk}(A|B, C), \qquad (9)$$

and so on. Accordingly, the generalized composition law becomes

$$\begin{aligned} S_q(A, B, C) &= S_q(B, C) + S_q(A|B, C) + (1-q) S_q(B, C) S_q(A|B, C) \\ &= S_q(C) + S_q(B|C) + S_q(A|B, C) \\ &\quad + (1-q) [S_q(C) S_q(B|C) + S_q(B|C) S_q(A|B, C) + S_q(A|B, C) S_q(C) \\ &\quad + (1-q) S_q(C) S_q(B|C) S_q(A|B, C)]. \end{aligned} \qquad (10)$$

Therefore, we obtain

$$S_q(B|C) = \frac{S_q(A, B, C) - [S_q(C) + S_q(A|B, C) + (1-q)S_q(C)S_q(A|B, C)]}{1 + (1-q)[S_q(C) + S_q(A|B, C) + (1-q)S_q(C)S_q(A|B, C)]}, \quad (11)$$

for example. It is of interest to observe that the system $A$ plays only an auxiliary role in this equation since $S_q(B|C)$ does not directly contain information on $A$. This discussion can be generalized to an arbitrary multipartite system in an obvious way.

Closing this section, we wish to mention that classical nonadditive information theory has recently been applied to the problem of source coding in Ref. [13].

### III. QUANTUM NONADDITIVE CONDITIONAL ENTROPY

Equation (7) is assumed to remain form invariant under its quantum mechanical generalization, that is,

$$S_q(B|A) = \frac{S_q(A, B) - S_q(A)}{1 + (1-q)S_q(A)}, \quad (12)$$

where $S_q(A, B) = S_q[\hat{\rho}(A, B)]$ and $S_q(A) = S_q[\hat{\rho}(A)]$ with $\hat{\rho}(A)$ the marginal density matrix given by the partial trace, $\hat{\rho}(A) = \text{Tr}_B \, \hat{\rho}(A, B)$.

In classical theory, the nonadditive conditional entropy is always nonnegative, whereas it can be negative in quantum theory, in general. An important point arising here is that occurrence of negative values is actually a signature of quantum entanglement. Consider a classically correlated state, or a separable state, of $(A, B)$

$$\hat{\rho}(A, B) = \sum_\lambda w_\lambda \, \hat{\rho}_\lambda(A) \otimes \hat{\rho}_\lambda(B), \tag{13}$$

where $w_\lambda \in [0, 1]$ with $\sum_\lambda w_\lambda = 1$. This state is a non-product state but is known to admit locally realistic hidden-variable models [14]. Let us write $\hat{\rho}_\lambda(A)$ and $\hat{\rho}_\lambda(B)$ in the following forms:

$$\hat{\rho}_\lambda(A) = \sum_a r_\lambda(a) |a\rangle\langle a|, \qquad \hat{\rho}_\lambda(B) = \sum_a s_\lambda(b) |b\rangle\langle b|, \tag{14}$$

where $|a\rangle$ and $|b\rangle$ are the orthonormal bases of $A$ and $B$, respectively, and $r_\lambda(a), s_\lambda(b) \in [0, 1]$ with $\sum_a r_\lambda(a) = \sum_b s_\lambda(b) = 1$. Then, the nonadditive quantum conditional entropy in Eq. (12) is calculated to be

$$S_q(B|A) = \frac{\sum_a \left[\sum_\lambda w_\lambda r_\lambda(a)\right]^q S_q(B|a)}{\sum_a \left[\sum_\lambda w_\lambda r_\lambda(a)\right]^q}, \qquad (15)$$

where

$$S_q(B|a) = \frac{1}{1-q}\left\{\sum_b \left[\pi(b|a)\right]^q - 1\right\}, \qquad (16)$$

$$\pi(b|a) = \frac{\sum_\lambda w_\lambda r_\lambda(a) s_\lambda(b)}{\sum_\lambda w_\lambda r_\lambda(a)}. \qquad (17)$$

Equation (15) is to be compared with Eq. (4). $\pi(b|a)$ in Eq. (17) has the same properties as the classical conditional probability distribution does: $\pi(b|a) \in [0, 1]$, $\sum_b \pi(b|a) = 1$. Therefore, $S_q(B|A)$ in Eq. (15) is nonnegative for any classically correlated states. In other words, negative values of the nonadditive conditional entropy indicate existence of nonclassical correlation, i.e., quantum entanglement. In recent works [5,6], this point has been discussed in detail for a class of the density matrices of a $2 \times 2$ system (i.e., a bipartite spin-1/2 system). In particular, the strongest limitation [15] on separability of a parametrized form of the Werner-Popescu state [14,16] has been obtained using the nonadditive conditional entropy [5]. Also, it has been shown how the nonaddditive conditional entropy is superior to the conditional von Neumann entropy, $S(B|A) = \lim_{q \to 1} S_q(B|A)$, and to the Bell inequality for constraining validity of local realism.

## IV. QUANTUM ENTANGLEMENT IN A CLASS OF STATES OF MULTIPARTITE SYSTEMS

As the simplest multipartite generalization of the previous discussion about a bipartite spin-1/2 system, first let us consider a tripartite spin-1/2 system. A parametrized form of the Werner-Popescu-like state of such a system is given by

$$\hat{\rho}(A, B, C) = \frac{1-x}{8} \hat{I}_2(A) \otimes \hat{I}_2(B) \otimes \hat{I}_2(C) + x |\Psi_2^{(3)}\rangle\langle\Psi_2^{(3)}| \quad (x \in [0, 1]), \qquad (18)$$

where $\hat{I}_2$ denotes the $2 \times 2$ unit matrix and

$$|\Psi_2^{(3)}\rangle = \frac{1}{\sqrt{2}} \left( |0\rangle_A \otimes |0\rangle_B \otimes |0\rangle_C + |1\rangle_A \otimes |1\rangle_B \otimes |1\rangle_C \right). \qquad (19)$$

Since the systems, $A$, $B$, and $C$, appear symmetrically, there are essentially two different kinds of the marginal density matrices:

$$\hat{\rho}(B, C) = \text{Tr}_A \, \hat{\rho}(A, B, C)$$

$$= \frac{1-x}{4} \hat{I}_2(B) \otimes \hat{I}_2(C) + \frac{x}{2} \left( |0\rangle_{BB}\langle 0| \otimes |0\rangle_{CC}\langle 0| + |1\rangle_{BB}\langle 1| \otimes |1\rangle_{CC}\langle 1| \right), \quad (20)$$

$$\hat{\rho}(C) = \text{Tr}_{A,B}\, \hat{\rho}(A, B, C)$$

$$= \frac{1}{2} \hat{I}_2(C). \tag{21}$$

The eigenvalues of $\hat{\rho}(A, B, C)$, $\hat{\rho}(B, C)$, and $\hat{\rho}(C)$ are respectively given by

$$\hat{\rho}(A, B, C): \quad \frac{1-x}{8} \text{ (7-fold degenerate)}, \quad \frac{1+7x}{8}, \tag{22}$$

$$\hat{\rho}(B, C): \quad \frac{1-x}{4} \text{ (doubly degenerate)}, \quad \frac{1+x}{4} \text{ (doubly degenerate)}, \tag{23}$$

$$\hat{\rho}(C): \quad \frac{1}{2} \text{ (doubly degenerate)}. \tag{24}$$

Therefore, the two independent nonadditive conditional entropies are calculated to be

$$S_q(A, B|C) = \frac{S_q(A, B, C) - S_q(C)}{1 + (1-q) S_q(C)}$$

$$= \frac{1}{1-q} \left[ \frac{7\left(\frac{1-x}{8}\right)^q + \left(\frac{1+7x}{8}\right)^q}{2\left(\frac{1}{2}\right)^q} - 1 \right], \tag{25}$$

$$S_q(A|B, C) = \frac{S_q(A, B, C) - S_q(B, C)}{1 + (1-q)S_q(B, C)}$$

$$= \frac{1}{1-q}\left[\frac{7\left(\frac{1-x}{8}\right)^q + \left(\frac{1+7x}{8}\right)^q}{2\left(\frac{1-x}{4}\right)^q + 2\left(\frac{1+x}{4}\right)^q} - 1\right]. \tag{26}$$

The zeros of Eqs. (25) and (26) are different each other. What we concern here is the value of $x$ as a function of $q$ on the border at which the nonadditive conditional entropy vanishes. It can be shown numerically that both $S_q(A, B|C) = 0$ and $S_q(A|B, C) = 0$ yield monotonically decreasing $x$ with respect to $q$. In particular, the latter is found to yield the following stronger limitation on the regime in which local realism holds:

$$0 \leq x < 1/5, \tag{27}$$

which is obtained in the limit $q \to \infty$. [Actually, the condition in Eq. (27) is necessary and sufficient for separability of $\hat{\rho}(A, B, C)$ in Eq. (18). See the general discussion below.] This suggests that what to be examined is the nonadditive conditional entropy of the form

$$S_q(A_1|A_2,A_3,\text{L},A_n) = \frac{S_q(A_1,A_2,\text{L},A_n) - S_q(A_2,A_3,\text{L},A_n)}{1+(1-q)S_q(A_2,A_3,\text{L},A_n)}, \quad (28)$$

in general. It should be noted that, *in the asymptotic evaluation of this quantity in the limit* $q \to \infty$, *it is sufficient to consider the largest eigenvalues of* $\hat{\rho}(A_1,A_2,\text{L},A_n)$ *and* $\hat{\rho}(A_2,A_3,\text{L},A_n)$, since $S_q(A_1|A_2,A_3,\text{L},A_n)$ can also be expressed as follows:

$$S_q(A_1|A_2,A_3,\text{L},A_n) = \frac{1}{1-q}\left[\frac{\text{Tr}\,\hat{\rho}^q(A_1,A_2,\text{L},A_n)}{\text{Tr}\,\hat{\rho}^q(A_2,A_3,\text{L},A_n)} - 1\right]. \quad (29)$$

Now, we consider the problem of separability of the parametrized Werner-Popescu-like state of an $N^n$-system (i.e., an $n$-partite $N$-level system). The density matrix of this state is written as follows:

$$\hat{\rho}(A_1,A_2,\text{L},A_n) = \frac{1-x}{N^n}\hat{I}_N(A_1) \otimes \hat{I}_N(A_2) \otimes \text{L} \otimes \hat{I}_N(A_n)$$
$$+ x|\Psi_N^{(n)}\rangle\langle\Psi_N^{(n)}| \quad (x \in [0,1]), \quad (30)$$

where $\hat{I}_N$ is the $N \times N$ unit matrix and $|\Psi_N^{(n)}\rangle$ is given by

$$|\Psi_N^{(n)}\rangle = \frac{1}{\sqrt{N}}\sum_{k=0}^{N-1} |k\rangle_{A_1} \otimes |k\rangle_{A_2} \otimes \text{L} \otimes |k\rangle_{A_n}. \quad (31)$$

The marginal density matrix of interest is

$$\hat{\rho}(A_2, A_3, \mathrm{L}, A_n) = \mathrm{Tr}_{A_1} \hat{\rho}(A_1, A_2, A_3, \mathrm{L}, A_n)$$

$$= \frac{1-x}{N^{n-1}} \hat{I}_N(A_2) \otimes \hat{I}_N(A_3) \otimes \mathrm{L} \otimes \hat{I}_N(A_n)$$

$$+ \frac{x}{N} \sum_{k=0}^{N-1} |k\rangle_{A_2\,A_2}\langle k| \otimes |k\rangle_{A_3\,A_3}\langle k| \otimes \mathrm{L} \otimes |k\rangle_{A_n\,A_n}\langle k|. \qquad (32)$$

It can be found that the eigenvalues of $\hat{\rho}(A_1, A_2, \mathrm{L}, A_n)$ and $\hat{\rho}(A_2, A_3, \mathrm{L}, A_n)$ are respectively given by

$$\hat{\rho}(A_1, A_2, \mathrm{L}, A_n): \quad \frac{1-x}{N^n} \; [(N^n-1)\text{-fold degenerate}], \quad \frac{1+(N^n-1)x}{N^n}, \qquad (33)$$

$$\hat{\rho}(A_2, A_3, \mathrm{L}, A_n): \quad \frac{1-x}{N^{n-1}} \; [(N^{n-1}-N)\text{-fold degenerate}],$$

$$\frac{1+(N^{n-2}-1)x}{N^{n-1}} \; (N\text{-fold degenerate}). \qquad (34)$$

Using these eigenvalues, the nonadditive conditional entropy in Eq. (28) [or Eq. (29)] is calculated to be

$$S_q(A_1 | A_2, A_3, \text{L}, A_n)$$

$$= \frac{1}{1-q}\left\{\frac{(N^n-1)\left(\frac{1-x}{N^n}\right)^q + \left[\frac{1+(N^n-1)x}{N^n}\right]^q}{(N^{n-1}-N)\left(\frac{1-x}{N^{n-1}}\right)^q + N\left[\frac{1+(N^{n-2}-1)x}{N^{n-1}}\right]^q} - 1\right\}. \quad (35)$$

For fixed $N$ and $n$, the value of $x$ satisfying $S_q(A_1 | A_2, A_3, \text{L}, A_n) = 0$ monotonically decreases with respect to $q$, as in the tripartite spin-1/2 system discussed previously. Thus, in the limit $q \to \infty$, evaluating the eigenvalues of $\hat{\rho}(A_1, A_2, \text{L}, A_n)$ and $\hat{\rho}(A_2, A_3, \text{L}, A_n)$, we find that the state in Eq. (30) is separable if

$$0 \leq x < \frac{1}{1+N^{n-1}}. \quad (36)$$

It has recently been shown using algebraic methods [17,18] that Eq. (36) is actually the necessary and sufficient condition. This indicates how the generalized nonadditive information theoretic approach sheds new light on characterizing quantum entanglement in a class of the Werner-Popescu-like states of multipartite systems.

## V. CONCLUSION

We have developed a basis for nonadditive generalization of the ordinary framework of quantum information. We have examined its potential utility for characterizing quantum entanglement in multipartite systems. In particular, we have considered a class of mixed states, which are generalizations of the parametrized form of the Werner-Popescu state. We have shown that the strongest limitation on validity of local realism in this class can be obtained in a novel manner.

**Acknowledgments**

The author would like to thank Professors. A. Mann, A. Peres, and M. Revzen for informative discussions. He also acknowledges the hospitality of Technion-Israel Institute of Technology extended to him. This work was supported in part by the Grant-in-Aid for Scientific Research of Japan Society for the Promotion of Science.